\documentstyle[12pt,epsf]{article}

\def\lsim{\mathrel{\raise.2ex\hbox{$<$}\hskip-.8em\lower.9ex\hbox{$\sim$}}}
\def\gsim{\mathrel{\raise.2ex\hbox{$>$}\hskip-.8em\lower.9ex\hbox{$\sim$}}}

\begin{document}

\thispagestyle{empty}

\font\fortssbx=cmssbx10 scaled \magstep2
\hbox to \hsize{
\hskip.35in \raise.1in\hbox{\fortssbx University of Wisconsin - Madison}
\hfill$\vcenter{\hbox{\bf MADPH-97-982}
            \hbox{\bf astro-ph/9702193}
            \hbox{February 1997}}$ }

\vspace{.5in}

\begin{center}
{\large\bf Neutrino Fluxes from Active Galaxies:\\
a Model-Independent Estimate}\\[4mm]
F. Halzen\\
{\it Department of Physics, University of Wisconsin, Madison, WI 53706}\\[2mm]
E. Zas\\
{\it Dpto.\ de F\'\i sica de Part\'\i culas, Universidad de Santiago, E-15706 Santiago,
Spain}
\end{center}

\vspace{.25in}

{\small\narrower
There are tantalizing hints that jets, powered by supermassive black holes at
the center of active galaxies, are true cosmic proton accelerators. They
produce photons of TeV energy, possible higher, and may be the enigmatic source
of the highest energy cosmic rays. Photoproduction of neutral pions by
accelerated protons on UV light is the source of the highest energy photons, in
which most of the bolometric luminosity of the galaxy may be emitted. The case
that proton beams power active galaxies is, however, far from conclusive.
Neutrinos from the decay of charged pions represent an uncontrovertible
signature for the proton induced cascades. We show that their flux can be
estimated by model-independent methods, based on dimensional analysis and
textbook particle physics. Our calculations also demonstrate why different
models for the proton blazar yield very similar results for the neutrino flux,
consistent with the ones obtained here.\par}

\vspace{.2in}

\def\large{\normalsize}
\def\Large{\normalsize}
\renewcommand{\thesection}{\arabic{section}.}

\section*{Introduction}

In recent years cosmic ray experiments have revealed the existence of cosmic
particles with energies in excess of 10$^{20}$~eV. Incredibly, we have no clue
where they come from and how they have been accelerated to this
energy\cite{watson}. The highest energy cosmic rays are, almost certainly, of
extra-galactic origin. Searching the sky beyond our galaxy, the nuclei of
active galaxies (AGN) stand out as the most likely sites of magnetic fields
which are sufficiently strong and expansive to accelerate particles to joules
of energy. The idea is rather compelling because AGN are also the source of the
highest energy photons, detected with air Cherenkov telescopes\cite{whipple}.

AGN are the brightest sources in the Universe. Their engines must not only be
powerful, but extremely compact because their high energy luminosities are
observed to flare by over an order of magnitude over time periods as short as a
day\cite{variability}. Only sites in the vicinity of black holes, a billion
times more massive than our sun, can possibly satisfy the constraints of the
problem. Highly relativistic and confined jets of particles are a common
feature of these objects. It is anticipated that beams, accelerated near the
black hole, are dumped on the radiation in the galaxy which consists of mostly
thermal photons with densities of order 10$^{14}$/cm$^3$. The multi-wavelength
spectrum, from radio waves to TeV gamma rays, is produced in the interactions
of the accelerated particles with the magnetic fields and ambient light in the
galaxy. In the more conventional electron models, the highest energy photons
are produced by Compton scattering of accelerated electrons on thermal UV
photons which are scattered from 10~eV up to TeV energy\cite{dermer}. The
energetic gamma rays will subsequently lose energy by electron pair production
in photon-photon interactions with the radiation field of the jet or the
galactic disk. An electromagnetic cascade is thus initiated which, via pair
production on the magnetic field and photon-photon interactions, determines the
emerging gamma-ray spectrum at lower energies. The lower energy photons,
observed by conventional astronomical techniques, are, as a result of the
cascade process, several generations removed from the primary high energy
beams.

The EGRET instrument on the Compton Gamma Ray Observatory has detected high
energy gamma-ray emission, in the range 20 MeV--30 GeV, from over 100
sources\cite{EGRET}. Of these sources 16 have been tentatively, and 42 solidly
identified with radio counterparts. All belong to the ``blazar" subclass,
mostly Flat Spectrum Radio Quasars, while the rest are BL-Lac
objects\cite{Mattox}. In a unified scheme of AGN, they correspond to Radio Loud
AGN viewed from a position illuminated by the cone of a relativistic
jet\cite{padovani}. Moreover of the five TeV gamma-ray emitters identified by
the air Cherenkov technique, three are extra-galactic and are also nearby
BL-Lac objects\cite{whipple}. The data therefore strongly suggests that the
highest energy photons originate in jets beamed to the observer. Several of the
sources observed by EGRET have shown strong variability, by a factor of 2 or so
over a time scale of several days\cite{variability}. Time variability is more
spectacular at higher energies. On May 7, 1996 the Whipple telescope observed
an increase of the TeV-emission from the blazar Markarian 421 by a factor 2 in
1 hour reaching, eventually, a value 50 times larger than the steady flux. At
this point the telescope registered 6 times more photons from the Markarian
blazar, more distant by a factor $10^5$, than from the Crab supernova
remnant\cite{421flare}.

Does pion photoproduction by accelerated protons play a central role in blazar
jets? This question has been extensively debated in recent years\cite{pic}. If
protons are accelerated along with electrons, they will acquire higher
energies, reaching PeV--EeV energy because of reduced energy losses. High
energy photons result from proton-induced photoproduction of neutral pions on
the ubiquitous UV thermal background. Accelerated protons thus initiate a
cascade which dictates the features of the spectrum at lower
energy\cite{biermann}. From a theorist's point of view the proton blazar has
attractive features. Protons, unlike electrons, efficiently transfer energy
from the black hole in the presence of the high magnetic fields required to
explain the confinement of the jets\cite{mannheimB}. Protons provide a
``natural'' mechanism for energy transfer from the central engine over
distances as large as 1~parsec, as well as for the observed heating of the
dusty disk over distances of several hundred parsecs\cite{biermann}. More to
the point, the issue of proton acceleration can be settled experimentally
because the proton blazar is a source of high energy protons and neutrinos, not
just gamma rays\cite{PR}.

Weakly interacting neutrinos can, unlike high energy gamma-rays and high energy
cosmic rays, reach us from more distant and much more powerful AGN. It is
likely that absorption effects explain why Markarian 421, the closest blazar on
the EGRET list at a distance of $\sim$~150~Mpc , produces the most prominent
TeV signal. Although the closest, it is one of the weakest; the reason that it
is detected whereas other, more distant, but more powerful, AGN are not, must
be that the TeV gamma rays suffer absorption in intergalactic space by
interaction with background infra-red light\cite{salomon}. This most likely
provides the explanation why much more powerful quasars with significant high
energy components such as 3C279 at a redshift of 0.54 have not been identified
as TeV sources.
Undoubtedly, part of the TeV flux is also absorbed on the infrared light in the
source; we will return to this further on.

\section{Modelling of Blazar Jets}

First order Fermi acceleration offers a very attractive model for acceleration
in jets, providing, on average, the right power and spectral shape. A cosmic
accelerator in which the dominant mechanism is first order diffusive shock
acceleration, will indeed produce a spectrum
\[
dN/dE\propto E^{-\gamma} \,, \label{eq:dN/dE}
\]
with $\gamma \sim 2 +\epsilon$, where $\epsilon$ is a small number. For strong
ultra-relativistic shocks it can be negative ($\sim -0.3$ ). Confronted with
the challenge of explaining a relatively flat multi-wavelength photon emission
spectrum which extends to TeV energy, models have converged on the blazar
blueprint shown in Fig.~1. Particles are accelerated by Fermi shocks in bunches
of matter travelling along the jet with a bulk Lorentz factor of order $\gamma
\sim 10$. Ultra-relativistic beaming with this Lorentz factor provides the
natural interpretation of the observed superluminal speeds of radio structures
in the jet\cite{rees}. In order to accommodate bursts lasting a day in the
observer's frame, the bunch size must be of order $\Gamma c \Delta t \sim
10^{-2}$~parsecs. Here $\Gamma$ is the Doppler factor, which for observation
angles close to the jet direction is of the same order as the Lorentz
factor~\cite{padovani}. These bunches are, in fact, more like sheets, thinner
than the jet's width of roughly 1~parsec. The observed radiation at all
wavelengths is produced by the interaction of the accelerated particles in the
sheets with the ambient radiation in the AGN, which has a significant component
concentrated in the so-called ``UV-bump\rlap".

\begin{figure}[t]
\centering
\epsfxsize=3in\hspace{0in}\epsffile{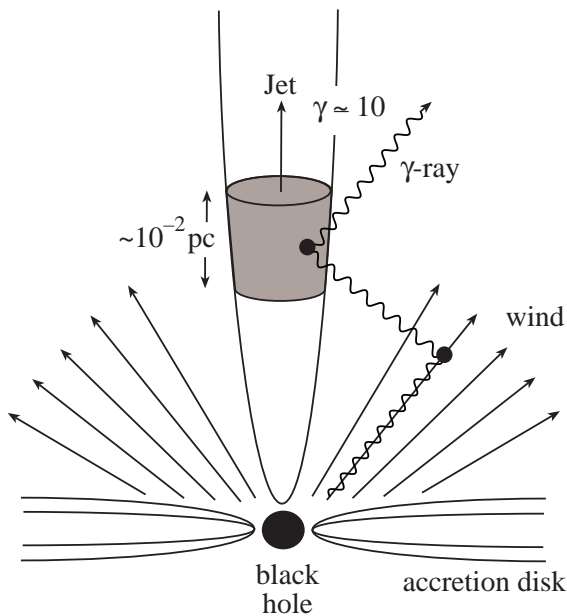}

\caption{Possible blueprint for the production of high energy photons and
neutrinos near the super-massive black hole powering an AGN. Particles,
accelerated in sheet like bunches moving along the jet, interact with photons
radiated by the accretion disk or produced by the interaction of the
accelerated particles with the magnetic field of the jet.}
\end{figure}

In electron models the multi-wavelength spectrum consists of  three components:
synchrotron radiation produced by the electron beam on the $B$-field in the
jet, synchrotron photons Compton scattered to high energy by the electron beam
and, finally, UV photons Compton scattered by the electron beam to produce the
highest energy photons in the spectrum\cite{dermer}. The seed photon field can
be either external, e.g.\ radiated off the accretion disk, or result from the
synchrotron radiation of the electrons in the jet, so-called
synchrotron-self-Compton models. The picture has a variety of problems. In
order to reproduce the observed high energy luminosity, the accelerating
bunches have to be positioned very close to the black hole. The photon target
density is otherwise insufficient for inverse Compton scattering to produce the
observed flux. This is a balancing act, because the same dense target will
efficiently absorb the high energy photons by $\gamma\gamma$ collisions. The
balance is difficult to arrange, especially in light of observations showing
that the high energy photon flux extends beyond TeV energy\cite{whipple}. The
natural cutoff occurs in the 10--100~GeV region\cite{dermer}. Finally, in order
to prevent the electrons from losing too much energy before producing the high
energy photons, the magnetic field in the jet has to be artificially adjusted
to less than 10\% of what is expected from equipartition with the radiation
density.

For these, and the more general reasons already mentioned in the introduction,
the proton blazar has been developed. In this model protons as well as
electrons are accelerated. Because of reduced energy loss, protons can produce
the high energy radiation further from the black hole. The more favorable
production-absorption balance far from the black hole makes it relatively easy
to extend the high energy photon spectrum above 10 TeV energy, even with bulk
Lorentz factors that are significantly smaller than in the inverse Compton
models. Two recent incarnations of the proton blazar illustrate that these
models can also describe the multi-wavelength spectrum of the
AGN\cite{mannheim,protheroe}. Because the seed density of photons is still much
higher than that of target protons, the high energy cascade is initiated by the
photoproduction of neutral pions by accelerated protons on ambient light via
the $\Delta$ resonance. The protons collide either with synchrotron photons
produced by electrons\cite{mannheim}, or with the photons radiated off the
accretion disk\cite{protheroe}, as shown in Fig.~1.

\section{The Neutrino Flux from Blazar Jets}

Model-independent evidence that AGN are indeed cosmic proton accelerators can
be obtained by observing high energy neutrinos from the decay of charged pions,
photoproduced on the $\Delta$ resonance along with the neutral ones. The
expected neutrino flux can be estimated in six easy steps.

\begin{enumerate}

\item

The size of the accelerator $R$ is determined by the duration, of order 1 day,
over which the high energy radiation is emitted:
\begin{equation}
R=\Gamma t c = 10^{-2}\mbox{ parsecs for }t = 1\rm\ day.
\end{equation}

\item

The magnitude of the $B$-field can be calculated from equipartition with the
electrons, whose energy density is measured experimentally:
\begin{equation}
{B^2 \over 2 \mu_0}  = \rho\rm (electrons) \,{\sim} \, 1\ erg/cm^3.
\end{equation}
This yields a value for the magnetic field of 5~Gauss. A similar value is
obtained by scaling $B$-fields in the jets of Fannaroff-Riley type II galaxies
at kiloparsec distances, to the Markarian 421 luminosity, and to transverse
distances in the milliparsec range\cite{mannheimB}.

\item

In shock acceleration the gain in energy occurs gradually as a particle near
the shock scatters back and forth across the front gaining energy with each
transit. The proton energy is limited by the lifetime of the accelerator and
the maximum size of the emitting region, $R$\cite{PR}
\begin{equation}
E  < K Z e B R c\,.
\label{eq:Emax}
\end{equation}
Here $Ze$ is the charge of the particle being
accelerated and $B$
the ambient magnetic field. The upper limit basically follows from dimensional
analysis. It can also be derived from the simple requirement that the
gyroradius of the accelerated particles must be contained within the
accelerating region $R$. The numerical constant $K\sim 0.1$ depends on the
details of diffusion in the vicinity of the shock, which determine the
efficiency by which power in the shock is converted into acceleration of
particles. In some cases it can reach values close to 1. The maximum energy
reached is
\[
E_{\rm max} = e B R c = 5 \times 10^{19} \rm\ eV
\]
for $B = 5$~Gauss and  $R = 0.02$~parsecs. We here assumed that the boost of
the energy in the observer's frame approximately compensates for the efficiency
factor, i.e.\ $K \Gamma \sim 1$.

The neutrino energy is lower by two factors which take into account i) the
average momentum carried by the secondary pions relative to the parent proton
($\left< x_F\right> \simeq 0.2$) and ii) the average energy carried by the
neutrino in the decay chain $\pi^+ \rightarrow \nu_\mu \mu^+ \rightarrow  e^+
\nu_e \bar{\nu}_\mu$, which is roughly 1/4 of the pion energy because equal
amounts of energy are carried by the four leptons. The maximum neutrino energy
is
\begin{equation}
E_{\nu\,\rm max} = E_{\rm max} \left< x_F\right> {1\over4} \simeq 10^{18}\rm\,
eV\,,
\end{equation}
i.e.\ neutrinos reach energies of $10^3$~PeV.

\item

The neutrino spectrum can now be calculated from the observed gamma ray
luminosity. We recall that approximately equal amounts of energy are carried by
the four leptons that result from the  decay chain $\pi^+ \rightarrow \nu_\mu
\mu^+ \rightarrow e^+ \nu_e \bar{\nu}_\mu$. In addition the cross sections for
the processes $p\gamma \rightarrow p \pi^0$ and  $p\gamma \rightarrow n \pi^+ $
at the $\Delta$ resonance are in the approximate ratio of $2:1$. Thus 3/4 of
the energy lost to photoproduction ends up in the electromagnetic cascade and
1/4 goes to neutrinos, which corresponds to a ratio of neutrino to gamma
luminosities ($L_\nu : L_\gamma$) of $1:3$. This ratio is somewhat reduced when
taking into account that some of the energy of the accelerated protons is lost
to direct pair production ($p +\gamma\rightarrow e^+ e^- p$):
\begin{equation}
L_\nu\, =\,\frac{3}{13}L_\gamma\,.
\end{equation}
In order to convert above relation into a neutrino spectrum we have to fix the
spectral index. We will assume that the target photon density spectrum is
described by a $E^{-(1+\alpha)}$ power law, where $\alpha$ is small for AGN
with flat spectra. The number of target photons above photoproduction threshold
grows when the proton energy $E_p$ is increased. If the protons are accelerated
to a  power law spectrum with spectral index $\gamma\,(=2+\epsilon)$, the
threshold effect implies that the spectral index of the secondary neutrino flux
is also a power law, but with an index flattened by $(1+\alpha)$ as a result of
the increase in target photons at resonance when the proton energy is
increased:
\begin{equation}
{dN_\nu\over dE_{\nu}} = {\cal N}
\left[
{ E_{\nu} \over E_{\nu\rm\,max} }
\right] ^{-(1+\epsilon-\alpha)}.
\end{equation}
For a standard non relativistic shock with $\epsilon = 0$ and a flat photon
target with $\alpha=0$, the neutrino spectrum will flatten by just one unit
giving $E{dN_\nu\over dE} \sim \rm constant$. From Eqs.~(5) and (6)
\begin{equation}
\int^{E_{\nu\rm\,max}} E{dN_\nu\over dE_{\nu}} dE_{\nu} \simeq
{\cal N} {E_{\nu\rm\,max}^2 \over 1-\epsilon+\alpha} \simeq {3\over13}L_\gamma
\,.
\end{equation}
The calculation is stable as long as $\epsilon - \alpha$ is smaller than 1
because the luminosity integral is not sensitive to the lower limit of the
integration.

\item

Assuming that the high energy $\gamma$ ray flux from Markarian 421 results from
cascading of the gamma ray luminosity produced by Fermi accelerated protons, we
obtain the neutrino flux from the measured value\cite{whipple} of $L_{\gamma}$
of $2 \times 10^{-10}$~TeV~cm$^{-2}$~s$^{-1}$:
\begin{equation}
{dN_\nu\over dE_{\nu}} = {3\over13}{L_\gamma\over E_{\nu\rm\,max}} \;{1- \epsilon +
\alpha \over E_{\nu}} \left[ { E_{\nu} \over E_{\nu\rm\,max} }
\right]^{-(\epsilon+\alpha)} \sim
{5 \times 10^{-17}\,{\rm cm^{-2}\,s^{-1}}\over E_{\nu}} \,,
\end{equation}
where the numerical estimate corresponds to $\alpha=\epsilon=0$ and the value
of $E_{\nu\rm\,\max}$ of Eq.~(4). This calculation reveals that for the small
values of $\epsilon$ and $\alpha$ anticipated, the neutrino flux is essentially
determined by the value for $E_{\nu\rm\,\max}$.

\item

In order to calculate the diffuse flux from the observed blazar distribution,
we note that the EGRET collaboration has constructed a luminosity function
covering the observation of the $\sim$20 most energetic blazars and estimated
the diffuse gamma ray luminosity\cite{chiang}. From the ratio of the diffuse
gamma ray flux and the flux of Markarian 421, we obtain that the effective
number of blazars with Markarian 421 flux is $\sim 130\rm~sr^{-1}$. The
diffuse neutrino flux is now simply estimated by multiplying the calculated
flux for Markarian 421 by this factor. A correction for the difference in
spectral indices of gamma ray and neutrino fluxes enhances the neutrino flux by
a factor of three. The flux corresponds to an energy regime well below the high
energy cut-off. The transition to the cutoff should be smooth because of the
superposition of the different redshifts and cut-off energies of the individual
blazars.
\end{enumerate}

This concludes our calculation. It illustrates how the proton blazar, unlike
the electron blazar, requires no large Doppler factors and no fine-tuning of
parameters. For the proton blazar, radiation and magnetic fields are in
equipartition, the maximum energy matches the $BR$ value expected from
dimensional analysis and, finally, the size of the bunches is similar to the
gyroradius of the highest energy protons. It is not a  challenge to increase
gamma ray energies well beyond the TeV energy range. Reasonable variations of
the values of magnetic field strength $B$, the efficiency parameter $K$ and the
Doppler boost factor $\Gamma$ may allow us to account for the highest energy
cosmic rays with $E\sim 3~10^{20}~$eV.

Also, our calculation demonstrates why models\cite{mannheim,protheroe} which
differ in many aspects, yield very similar predictions for the neutrino flux,
consistent with the ones obtained here\cite{hill}; this is illustrated in
Fig~2.

\begin{figure}
\centering
\epsfxsize=4in\hspace{0in}\epsffile{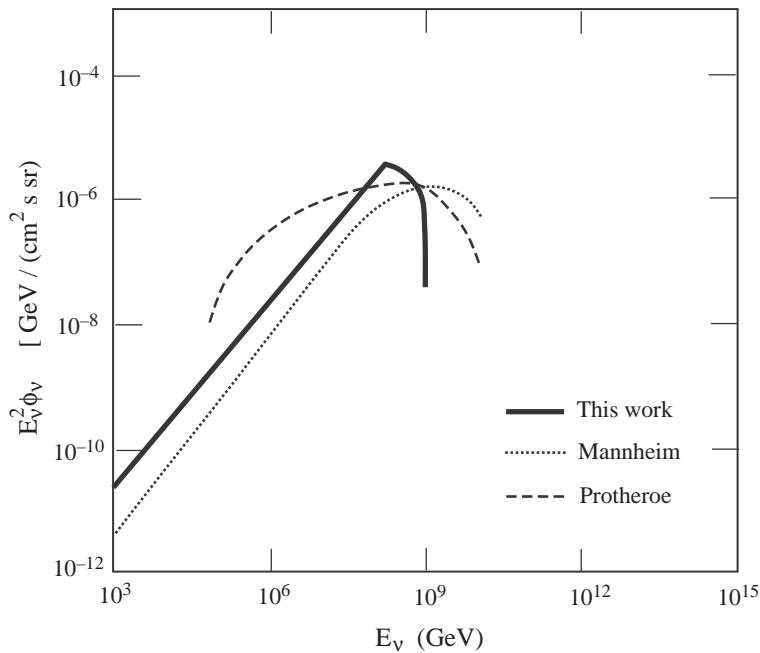}

\caption{Diffuse neutrino flux from blazars. The numerical result of Equation~(8) multiplied by $3 \times 130$~sr$^{-1}$ and corrected for redshift in the cutoff is compared to recent calculations \protect\cite{mannheim,protheroe}.}
\end{figure}

\section{The Cosmic Ray Argument}

Rather than scaling to the TeV gamma ray flux, we can use the cosmic ray flux
at ultra high energies to bracket expectations for the neutrino flux. Models
for proton acceleration in hot spots of Fanaroff-Riley type II galaxies can
explain the observed cosmic ray spectrum above $\sim 10^{18}$~eV\cite{rachen}.
This energy corresponds to the ``ankle" in the spectrum, where the observed
spectral index flattens from 3 to 2.7. The model requires an $E^{-2}$, or
flatter, injection spectrum which steepens above $10^{17}$~eV to the observed
$E^{-2.7}$ spectrum as a result of energy loss in the source, interactions with
the microwave background, and cosmological evolution\cite{rachen}. Because of
the strict limitations on the density of target photons at the acceleration
site, previously discussed, roughly similar neutrino and proton luminosities
are expected\cite{PR}. In order to understand this balance it is important to
realize that in astrophysical beam dumps the accelerator and production target
form a symbiotic system. Although larger target density may produce more
neutrinos, it also decelerates the protons producing them, in a delicate
acceleration-absorption balance. Equal cosmic ray and neutrino luminosity
implies:
\begin{equation}
\int dE_{\nu} (E_{\nu} \, dN_\nu/dE_{\nu}) \sim L_{\rm CR} \sim 10^{-9}\rm\ TeV\
cm^{-2}\, s^{-1} \, sr^{-1} \;.
\end{equation}
The bulk cosmic ray luminosity has been conservatively estimated by assuming
that it is due to an $E^{-2.7}$ spectrum above $\sim 10^{17}~$eV. This spectrum
has been normalized to the observed EeV cosmic rays. It is interesting to note
that this luminosity is a factor 25 below the measured diffuse gamma ray
luminosity from AGN\cite{chiang}. This is, within an order of magnitude, in
agreement with the relation of neutrino and gamma ray luminosities of Eq.~(5),
and, if anything, implies a conservative estimate of the neutrino flux.

Assuming a generic $E^{-2}$ neutrino spectrum, the equality of cosmic-ray and
neutrino luminosities implies:
\begin{equation}
E_{\nu}{dN_\nu\over dE_{\nu}} \sim {10^{-10}\over E_{\nu}\,(\rm TeV)} \, \rm
cm^{-2}\ s^{-1}\ sr^{-1} \;. \label{eq:flux}
\end{equation}
A not too different result is obtained by assuming equal numbers of neutrinos
and protons, rather than equal luminosities. It is clear that our estimate is
conservative because the proton flux reaching Earth has not been corrected for
absorption of protons in ambient matter in the source, or in the interstellar
medium.

\section{Event Rates in Underground Muon Neutrino Telescopes}

The probability to detect a TeV neutrino is roughly $10^{-6}$\cite{PR}. It is
easily computed from the requirement that, in order to be detected, the
neutrino has to interact within a distance of the detector which is shorter
than the range of the muon it has produced. In other words, in order for the
neutrino to be detected, the produced muon has to reach the detector.
Therefore,
\begin{equation}
P_{\nu\to\mu} \simeq {R_\mu\over \lambda_{\rm int}} \simeq A E_{\nu}^n \,,
\end{equation}
where $R_{\mu}$ is the muon range and $\lambda_{\rm int}$ the neutrino
interaction length. For energies below 1~TeV, where both the range and cross
section depend linearly on energy, $n=2$. Between TeV and PeV energies $n=0.8$
and $A=10^{-6}$, with $E$ in TeV units. For EeV energies $n=0.47$, $A =10^{-2}$
with $E$ in EeV.

We are now ready to compute the diffuse neutrino event rate by folding the
neutrino spectrum of Eq.~(8) with the detection probability of Eq.~(11). We
also multiply by 130~sr$^{-1}$ for the effective number of sources:
\begin{equation}
\phi^\nu = \int^{E_{\nu\,\rm max}} {dN_\nu\over dE_{\nu}} P_{\nu\to\mu}(E_{\nu}) dE_{\nu} \simeq
40\rm\ km^{-2}\,year^{-1}\,sr^{-1} \,.
\end{equation}
which implies a yield of two neutrinos every three days in a kilometer-scale detector, assuming only $2 \pi$ coverage.

The steeper, but lower luminosity, flux of Eq.~(10) predicts more events when
folded with Eq.~(11), about $150\rm\ km^{-2}\,year^{-1}\,sr^{-1}$ assuming that
the flux extends down to TeV energy. The result does not depend strongly on the
lower limit of the neutrino integral, it only drops by a factor of three if the
neutrino flux flattens below 100~TeV. We again conclude that a kilometer-scale
neutrino detector may be required\cite{halzen}. It is however important to
realize that, had we assumed a $E^{-1}$ spectrum, the resulting flux would have
scaled with the ratio of luminosities to about an order of magnitude below
Eq.~(12). The energy dependence of the detection efficiency of  underground
muon neutrino detectors is such that most of the events are detected in the
high (low) energy end for a $E^{-1}$ ($E^{-2}$) spectrum.

\section{Evidence for the Proton Blazar?}

Astronomy with protons becomes possible once their energy has reached a value
where their gyroradius in the microgauss galactic field exceeds the dimensions
of the galaxy. Provided intergalactic magnetic fields are not too strong,
protons with $10^{20}$~eV energy point at their sources with degree-accuracy.
At this energy their mean-free-path in the cosmic microwave background is
unfortunately reduced to only tens of megaparsecs. A clear window of
opportunity emerges: Are the directions of the cosmic rays with energy in
excess of  ${\sim} 5 \times 10^{19}$~eV correlated to the nearest AGN
(red-shift $z$ less than 0.02), which are known to be clustered in the
so-called ``super-galactic" plane? Although far from conclusive, there is some
evidence that such a correlation may exist\cite{stanev}. Lack of statistics at
the highest energies is a major problem. Future large aperture cosmic ray
detectors such as the new Utah HIRES air fluorescence detector and the Auger
giant air shower array will soon remedy this aspect of the
problem\cite{watson}.

We have already drawn attention to the 10~TeV maximum photon energy as the
demarkation line between the electron and proton blazars. The $\sim10$~GeV
cutoff in the inverse Compton model can be pushed to the TeV range in order to
accommodate the Whipple data on Markarian 421, but not beyond. Bringing the
accelerator closer to the black hole may yield photons in excess of 10~TeV
energy --- they have, however, no chance of escaping without energy loss on the
dense infrared background at the acceleration site. HEGRA has been monitoring
the 10 closest blazars, including Markarian 421, with its dual telescope
systems: the scintillator and the naked photomultiplier detector arrays. The 
announcement\cite{rhode} that their upper limit on the photon flux of 50~TeV
and above for the aggregate emission from the ten nearest blazars, may be a
signal, could provide the first compelling evidence that blazar jets are indeed
proton accelerators.

In summary, there are hints that active galaxies may be true particle
accelerators with proton beams dictating the features of the spectrum. With the
rapidly expanding Baikal and AMANDA detectors producing their first hints of
neutrino candidates\cite{domogatsky,hulth}, observation of neutrinos from AGN
would establish the production of pions and identify the acceleration of
protons as the origin of the highest energy photons. A definite answer may not
be known until these detectors reach kilometer size. Neutrino telescope
builders should take note that, although smaller neutrino fluxes are predicted
than in the generic AGN models of a few years ago\cite{stecker}, they are all
near PeV energy where the detection efficiency is increased and the atmospheric
neutrino background negligible. Because of the beaming of the jets, the
neutrinos have a flatter spectrum peaking near the $10^6$~TeV maximum energy.
The actual event rates are, in the end, not very different.

If confirmed, these models strongly favor the construction of neutrino
telescopes following a distributed architecture, with large spacings of the
optical modules and relatively high threshold\cite{halzen}. This also opens up
opportunities for alternative techniques such as the radio technique, or the
detection of horizontal air showers with giant air shower arrays\cite{parente}.
Optimists, on the other hand, can find reasons to anticipate the discovery of
AGN neutrinos with much smaller telescopes. With a sufficiently high proton
target density in the acceleration region, much larger fluxes of neutrinos may
be produced in a proton-proton cascade. The predicted fluxes are however
model-dependent\cite{protheroe}. It is also possible, even likely, that
accelerated protons which produce neutrinos do not escape the source, or escape
after significant energy loss. Such absorption effects increase the neutrino
flux relative to the observed high energy cosmic ray flux, also leading to
larger neutrino fluxes.

\section*{Acknowledgements}

We benefitted from discussion with Peter Biermann, Jose Juan Blanco-Pillado,
Manuel Drees, Charles Goebel, Karl Mannheim, Raymond Protheroe and Todor
Stanev. This work was supported in part by the University of Wisconsin Research
Committee with funds granted by the Wisconsin Alumni Research Foundation, in
part by the U.S.~Department of Energy under Grant No.~DE-FG02-95ER40896, in
part by the CICYT under contract AEN96-1773 and by the Xunta de Galicia under
contract XUGA 20604A96.

\end{document}